\pdfoutput=1
\documentclass[final]{siamltex}
\usepackage{epsfig}
\usepackage{amsmath,amssymb}

\newcommand{\sgn}{\ensuremath{\mathrm{sgn}}} 
\newcommand{\pd}[2]{\ensuremath{\frac{\partial{}#1}{\partial{}#2}}}

\newcommand{\intR}{\int_{-\infty}^{\infty}}
\newcommand{\zt}{\tilde{z}}

\title{Asymptotic dynamics of \\ attractive-repulsive swarms}

\author{Andrew J. Leverentz\footnotemark[1] \and Chad M. Topaz\footnotemark[2]\ \and Andrew J. Bernoff\footnotemark[1]}

\begin{document}

\maketitle

\renewcommand{\thefootnote}{\fnsymbol{footnote}}
\footnotetext[1]{Dept. of Mathematics, Harvey Mudd College, Claremont, CA 91711}
\footnotetext[2]{Dept. of Mathematics and Computer Science, Macalester College, St. Paul, MN 55105}
\renewcommand{\thefootnote}{\arabic{footnote}}

\begin{abstract}
We classify and predict the asymptotic dynamics of a class of swarming models. The model consists of a conservation equation in one dimension describing the movement of a population density field. The velocity is found by convolving the density with a kernel describing attractive-repulsive social interactions. The kernel's first moment and its limiting behavior at the origin determine whether the population asymptotically spreads, contracts, or reaches steady-state. For the spreading case, the dynamics approach those of the porous medium equation. The widening, compactly-supported population has edges that behave like traveling waves whose speed, density and slope we calculate. For the contracting case, the dynamics of the cumulative density approach those of Burgers' equation. We derive an analytical upper bound for the finite blow-up time after which the solution forms one or more $\delta$-functions.
\end{abstract}



\pagestyle{myheadings}
\thispagestyle{plain}
\markboth{A.J. LEVERENTZ, C.M. TOPAZ AND A.J. BERNOFF}{ASYMPTOTIC DYNAMICS OF SWARMS}

\section{Introduction}
\label{sec:introduction}

Biological aggregations such as fish schools, bird flocks, ungulate herds, and insect swarms have drawn considerable attention from mathematical modelers in recent years. These animal groups -- which for brevity we refer to simply as swarms -- have implications for ecological dynamics, human food supply availability, disease transmission, and, on the longest spatiotemporal scales, evolution  \cite{ogk2001,tk1998}. Increasingly, they serve as prototypes for the development of algorithms in robotics, engineering, and artificial intelligence \cite{bdt1999,p2005}. Furthermore, biological swarms are a rich and versatile source of pattern-forming behavior, taking on morphologies including vortices, advancing fronts, branched dendritic structures, and more exotic patterns~\cite{edl2007,pk1999}.

The emergent organization of swarms can be mediated by exogenous influences such as nutrients, light, or gravity, as well as by endogenous ones, namely social interactions between individuals. Since many species swarm even in the absence of meaningful external stimuli, one concludes that social interactions play a key role. The most important social forces are thought to be attraction, repulsion, and alignment \cite{ckjrf2002,edl2007,edll2006}. Attraction refers to the evolutionarily preprogrammed tendency of conspecific organisms to move towards each other, which offers benefits such as protection and mate choice, while repulsion refers to the tendency to move away, for instance, for collision avoidance \cite{ogk2001}. Attraction and repulsion are driven by the relative locations of organisms. In contrast, alignment refers to the tendency of an organism to match the speed and orientation of its neighbors.

As highlighted in \cite{edl2007}, the particular combination(s) of attraction, repulsion, and alignment that are included in a model strongly affect the classes of solutions observed. For example, \cite{edl2007,edll2006} have elucidated the importance of alignment in giving rise to diverse and exotic swarming patterns including pulses, breathers, and ripples. In contrast, models including attraction and repulsion as the only social forces have a history of several decades and an extensive literature, much of which is reviewed in \cite{mkbs2003}. These models typically give rise to groups that spread, contract, or reach equilibrium~\cite{mkbs2003,tbl2006,tb2004}. If organisms are self-driven in addition, milling and migrating groups may form \cite{dcbc2006,lrc2001,mk1999}. 

In mathematical descriptions of swarms, a common modeling assumption is that social interactions take place in a pairwise manner, and that the effect of multiple organisms on a given organism can be determined via a superposition. Consider a swarm with a sufficiently large population such that the group is well-described by a continuum density $\rho(\vec{x},t)$, as in \cite{cdmbc2007,mk1999,tbl2006,tb2004} and many others. Under the aforementioned modeling assumptions, social forces involve a convolution term of the form
\begin{equation}
\int f_s(\vec{x}-\vec{y}) \rho(\vec{y},t) \, d\vec{y} \equiv f_s * \rho.
\end{equation}
Here $f_s$ is a kernel describing the social influence of the population at location $\vec{y}$ on that at location $\vec{x}$. Not only does the choice of social forces included in a model play a key role (as mentioned above), but the particular shape of the social kernel $f_s$ used to model a given social force can have a crucial affect on the dynamics of the group. For instance, the particular shape of the attractive-repulsive kernel used in \cite{cdmbc2007,dcbc2006} determined whether groups collapsed into a dense group, formed a well-spaced vortex-like swarm, formed a ring-like structure, or took one of  several other morphologies. 

If modelers are without explicit biological measurements giving an idea of a particular organism's social kernel, they face a crucial question: in order to construct a model that gives the qualitatively correct swarming behavior, what kernel should be chosen? One might think that since the kernel is a function, it determines an infinite-dimensional parameter space, and so selecting a particular point in that space for one's model might be challenging. In practice, modelers typically choose a functional form that is presumed to be phenomenologically appropriate, for instance, a kernel $f_s$ that is exponentially decaying in space and has the correct near-field and far-field behavior. For a few examples, see Table 1 in \cite{mkbs2003}. Even with such constraints, models may contain many parameters. For instance, there are five parameters controlling the social interactions used in \cite{lrc2001}, and at least eleven in \cite{edll2006}.

In this paper we analyze a given class of swarming models with a general social interaction kernel, and we classify and predict the possible asymptotic dynamics. The class of models we consider is
\begin{subequations}
\label{eqnOfMotion}
\begin{gather}
\rho_t + (\rho v)_x = 0 \\
v = \int_{-\infty}^{\infty} f_s(x-y) \rho(y) \, dy \equiv f_s * \rho. \label{velocityEquation}
\end{gather}
\end{subequations}
This equation describes a conserved continuum density field $\rho(x,t)$ on the real line. The velocity $v(x,t)$ depends exclusively on social interactions by means of a convolution with a kernel $f_s$ describing attraction and repulsion. In this paper, we focus solely on attractive-repulsive interactions, and hence do not consider social forces with an intermediate neutral zone as in, \emph{e.g.,} \cite{glr1996}; nor do we consider alignment. This model is kinematic, as opposed to dynamic, in which case the velocity would obey a momentum equation. As reviewed in \cite{edll2006}, social forces take place when animals communicate, either directly by auditory, visual, olfactory, or tactile senses, or indirectly, as mediated by chemical, vibrational, or other sorts of signals. A given type of communication may be unidirectional, as with visual sensing, or omnidirectional, as with auditory and olfactory sensing. Many organisms can process a combination of different input signals, which results in communication that is effectively omnidirectional \cite{edll2006,pm2005}. For this reason, in our one-dimensional model we choose $f_s$ to describe antisymmetric social forces, that is, we assume that $f_s$ is an odd function to ensure that distinct organisms exert equal and opposite forces on each other. Within the framework of (\ref{eqnOfMotion}), when $\sgn(x) f_s(x) < 0$ then the effective social force is attractive, and when $\sgn(x) f_s(x) > 0$ it is repulsive. Swarming models of the form (\ref{eqnOfMotion}) have been studied in \cite{bl2007,bv2005,bv2006} for specific choices of $f_s$, and in a two-dimensional setting in \cite{tb2004}.

\begin{figure}[t] 
\centerline{\epsfig{file=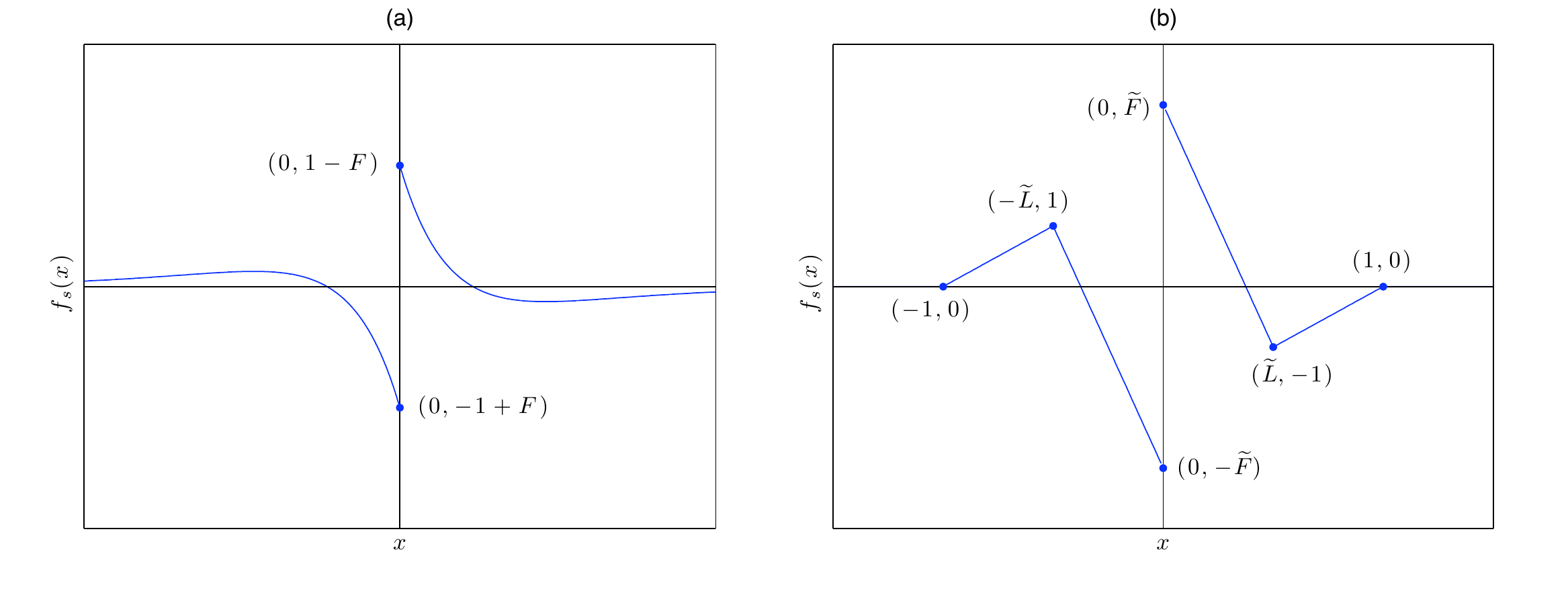, scale=0.58}}
\caption{Schematic depiction of $f_s$, the social interaction kernel in (\ref{eqnOfMotion}). (a) The Morse-type social force given by (\ref{morse}) for the case $F < 1$, $L > 1$. (b) The piecewise-linear social force given by (\ref{linearpotential}) for the case $\widetilde{F}>0$.}
\label{socialforces}
\end{figure}

A common choice for $f_s$ used in \emph{e.g.} \cite{dcbc2006,mkbs2003,tblt2008} and quite a few other studies is the Morse interaction force
\begin{equation}
f_s(x) = \sgn(x) \left[ -Fe^{-|x|/L} + e^{-|x|}\right].
\label{morse}
\end{equation}
Here, the first exponentially decaying term represents attraction with strength $F>0$ and characteristic length scale $L>0$. The second term, of opposite sign, describes repulsion. The problem has been nondimensionalized so that the repulsive strength and length scale are unity. Figure \ref{socialforces}(a) shows a schematic example of (\ref{morse}) for the case $F < 1$, $L > 1$. The Morse function is, in fact, a member of the more general class of kernels
\begin{equation}
f_s(x) = \sgn(x) \left[ -Fg(|x|/L) + g(|x|)\right],
\label{reproducing}
\end{equation}
where we scale the length and magnitude of $g$ such that it has first moment equal to $2$ and $g(0^+) = 1$. Here, $g(x)$ is some suitable function: it could be a Gaussian, a compactly supported function, or one of many other choices. We analyze both the Morse function (\ref{morse}) and the more general class (\ref{reproducing}) in this paper. However, our goal is to analyze (\ref{eqnOfMotion}) with as few assumptions on $f_s$ as possible, so we also consider cases more general than (\ref{morse}) and (\ref{reproducing}). 

We have already assumed $f_s$ is odd. We make three additional, relatively weak assumptions in order to facilitate our analysis. First, $f_s$ has a finite first moment. This assumption is consistent with the idea that organisms should not interact at very long length scales because their range of sensing is limited. Second, $f_s$ is continuous and piecewise differentiable everywhere except for a finite jump discontinuity at the origin. The biological intuition that supports this assumption is as follows: for a given organism, the effect of other organisms in the far-field should vary continuously with distance. Small changes in distance should induce small, continuous changes in the social force. However, since $f_s$ is odd, it is discontinuous at $x=0$ unless  $f_s(x) \to 0$ as $x \downarrow 0$. This is a degenerate case which we exclude here since we expect organisms in close proximity to have nonzero effects on each other. Note that this assumption implicitly excludes the case of so-called ``hard-core'' forces that blow up at $x=0$ \cite{mkbs2003}. Third, $f_s$ crosses $0$ for at most one value of $|x|$. We concentrate on the most biologically relevant case, when organisms are repelled at short distances (avoiding collision) and attracted at longer ranges (creating a tendency to form a swarm). This means that when two organisms are within sensing range of each other, they have a unique pairwise equilibrium distance. For completeness, we will also consider other cases captured within our modeling framework, namely some cases where organisms only repel (i.e.  $f_s \ge 0$ for $x>0$) or only attract (i.e.  $f_s \le 0$ for $x>0$), and briefly the unbiological case where there is attraction at short distances and repulsion at large distances.

Our main results are as follows. Eq. (\ref{eqnOfMotion}), with $f_s$ as described above, has three possible asymptotic behaviors. The population density profile can spread, blow up, or reach steady state. Via long-wave and short-wave analyses, we predict how the asymptotic dynamics depend on $f_s$. Specifically, the long-time behavior depends on two parameters which may be directly computed from $f_s$: the first moment and the limiting behavior at the origin. We perform numerical simulations of (\ref{eqnOfMotion}) to confirm these predictions for two example kernels. For the spreading case, the dynamics approach those of the porous medium equation. The widening, compactly-supported population has edges that behave like traveling waves whose speed, density and slope we calculate. For the contracting case, the dynamics of the cumulative density approach those of Burgers' equation. We derive an analytical upper bound for the finite blow-up time after which the solution forms one or more $\delta$-functions. The case of steady-state solutions is studied in \cite{bt2008}.

The remainder of this paper is organized as follows. In Section \ref{sec:predictions} we perform the long- and short-wave analyses of (\ref{eqnOfMotion}) to derive conditions on $f_s$ for the three possible asymptotic behaviors, and we confirm these predictions with numerical simulations. We also derive (local) equations describing the asymptotic dynamics. Section \ref{spreading} presents the spreading case in further detail, including an analysis of the traveling-wave-like behavior of the edge of the spreading group. Section \ref{sec:blowup} studies the blow-up case in more detail, including the analytical calculation bounding the finite blow-up time of the solution. We conclude in Section \ref{sec:conclusions}. At the end of this paper, there are two appendices. Appendix \ref{sec:moments} demonstrates conservation of mass and center of mass for (\ref{eqnOfMotion}). Appendix \ref{sec:numericalmethod} gives an overview of a particle-based numerical method we developed to simulate the model.

\section{Asymptotic behavior of solutions}
\label{sec:predictions}

\begin{figure}[ht] 
\centerline{\epsfig{file=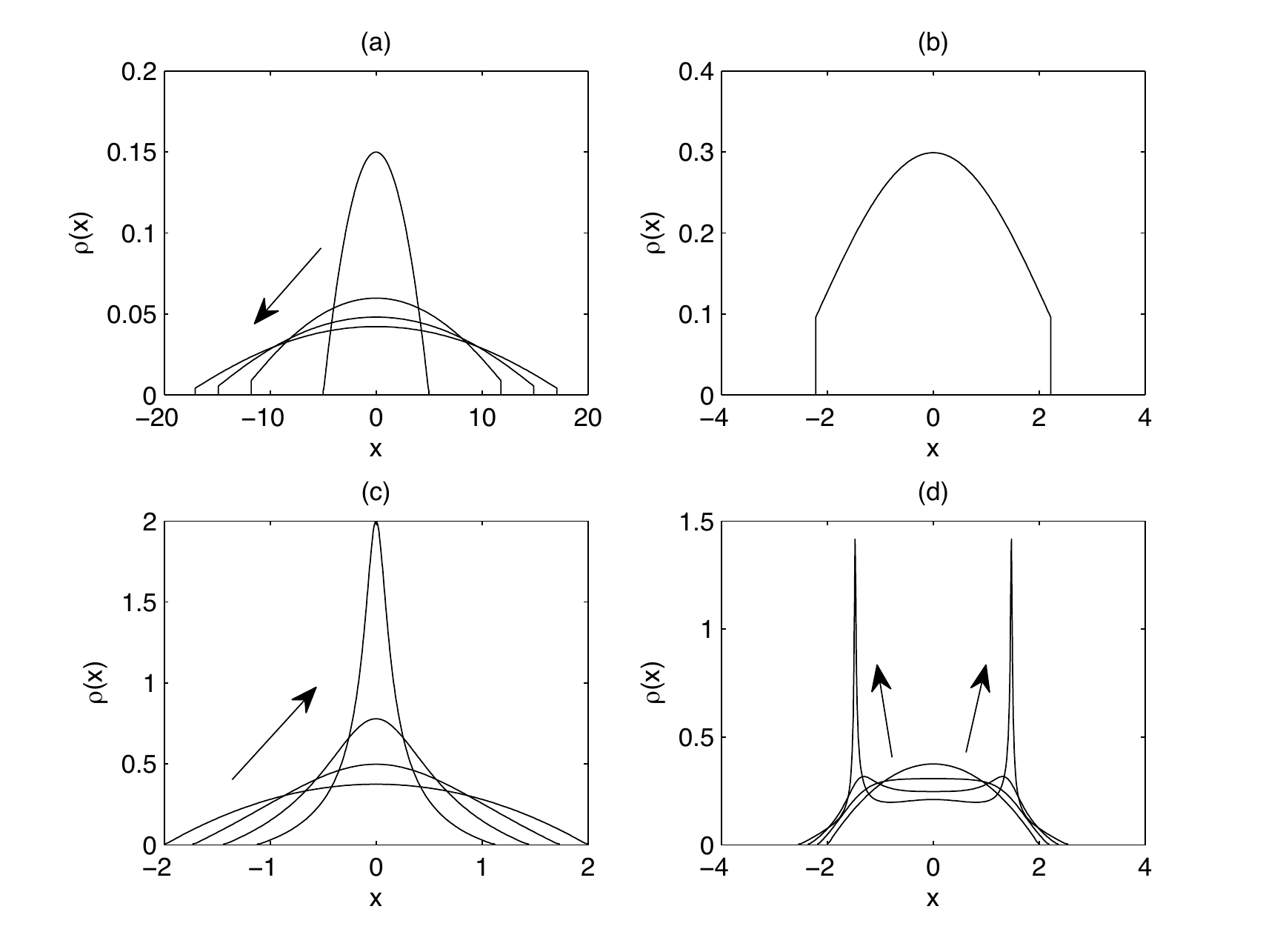, scale=0.8}}
\caption{Population density profiles governed by (\ref{eqnOfMotion}) with Morse-type social interactions (\ref{morse}). Arrows indicate the evolution of a profile over time. The asymptotic behavior of the model depends crucially on the choice of $F$, the relative strength of social attraction to social repulsion, and $L$, their relative characteristic length scales. (a) $F = 0.2$, $L = 2$. The compactly supported population eventually spreads to infinity. (b) $F = 0.4$, $L = 4$. The population reaches a compactly-supported steady state. (c) $F = 2$, $L = 2$. The density profile blows up into a single clump. (d) $F = 2$, $L = 0.5$. The density profile blows up by forming mutually repulsive clumps.  }
\label{snapshots}
\end{figure}

To demonstrate possible asymptotic behaviors of (\ref{eqnOfMotion}), we conduct numerical simulations using  the Morse-type social interaction (\ref{morse}) as an example. Our simulation takes place on an infinite domain and uses a particle-based numerical method we have developed, described in Appendix~\ref{sec:numericalmethod}. Simulations reveal three asymptotic behaviors, namely spreading, steady-state, and blow-up, as depicted in Figure \ref{snapshots}. Figure \ref{snapshots}(a) shows a spreading solution, corresponding to a population that disperses to infinity. The population density profiles are compactly supported, with a jump discontinuity at the edge. The profiles appear to be self-similar; we discuss this issue further in Sections  \ref{sec:shortwave} and  \ref{spreading}.  Figure \ref{snapshots}(b) shows a steady-state solution, corresponding to a localized aggregation of the population. Again, the population density drops discontinuously at the edge of the support. Steady states of (\ref{eqnOfMotion}) are analyzed in \cite{bt2008}. Figures \ref{snapshots}(c,d) show two cases of solutions where the density blows up, corresponding to populations with finite attraction at short distances. In the first case, the solution forms a single clump. In the second case,  the solutions form multiple, mutually-repelling clumps. These clumps are, in fact, $\delta$-functions, as we discuss in Section \ref{sec:blowup}. Our goal for the remainder of the present section is to derive conditions on a general social force $f_s$ to produce each of the aforementioned behaviors. To do this, we examine separately the long-wave and short-wave behavior of the system.

\subsection{Long-wave behavior}
\label{sec:longwave}
We first consider the evolution of initial conditions that are wide and slowly-varying. Specifically, assume that $\rho$ is initially long-wave, meaning $\widehat{\rho}$ is localized near wave number $k = 0$. We show that such initial conditions evolve, at least for a short time, according to the porous medium equation. To begin, let us define the Fourier transform of a function $h(x)$ as
\begin{equation}
\widehat{h}(k) = \mathcal{F}\{h\} = \int_{-\infty}^{\infty} h(x) e^{-ikx} \, dx.
\end{equation}
We apply the Convolution Theorem to (\ref{velocityEquation}) to write the Fourier transform of the velocity as
\begin{equation}
  \widehat{v}(k) = \mathcal{F}\{\rho * f_s\} = \widehat{\rho}(k) \widehat{f}_s(k).
\end{equation}
We next write $\widehat{f}_s(k)$ as a Taylor series
\begin{equation}
\label{eq:vhat}
  \widehat{v}(k) = \widehat{\rho}(k) \sum_{n=0}^{\infty} \frac{k^n}{n!} \widehat{f}_s^{(n)}(0).
\end{equation}
Then, we express the $n^{\textrm{th}}$ derivative of $\widehat{f}_s$ at $k=0$ in terms of the moments of $f_s$.  The $n^{\textrm{th}}$ moment of $f_s$ is
\begin{equation}
M_n[f_s] = \intR z^n \, f_s(z) \, dz.
\end{equation}
Then,
\begin{subequations}
\begin{eqnarray}
  \widehat{f}_s^{(n)}(0) &=& \left[ \frac{d^n}{dk^n} \intR f_s(x) e^{-ikx} \, dx \right]_{k=0} \\
    &=& \left[ \intR f_s(x) \frac{d^n}{dk^n} e^{-ikx} \, dx \right]_{k=0}  \\
    &=& (-i)^n \left[ \intR x^n f_s(x) e^{-ikx} \, dx \right]_{k=0} \\
    &=& (-i)^n \intR x^n f_s(x) \, dx \\
    &=& (-i)^n M_n[f_s]. \label{eq:momenthelp}
\end{eqnarray}
\end{subequations}
Substituting (\ref{eq:momenthelp}) into (\ref{eq:vhat}), we obtain
\begin{subequations}
\begin{eqnarray}
  \widehat{v}(k) &=& \sum_{n=0}^{\infty} \frac{(-1)^n}{n!} (ik)^n \widehat{\rho}(k) M_n[f_s] \\
  &=& \sum_{n=0}^{\infty} \frac{(-1)^n}{n!} \mathcal{F}\left\{ \frac{\partial^n \rho}{\partial x^n} \right\} M_n[f_s].
\end{eqnarray}
\end{subequations}
Since $f_s$ is antisymmetric, the even moments of $f_s$ vanish and we have
\begin{equation}
  \widehat{v}(k) = - \sum_{n=0}^{\infty} \frac{1}{(2n+1)!} \mathcal{F}\left\{ \frac{\partial^{2n+1} \rho}{\partial x^{2n+1}} \right\} M_{2n+1}[f_s].
  \label{vLongWaveSeries}
\end{equation}
or in physical space
\begin{equation}
  v(x)  = - \sum_{n=0}^{\infty} \frac{M_{2n+1}[f_s].}{(2n+1)!}  \frac{\partial^{2n+1} \rho}{\partial x^{2n+1}}
  \label{vLongWaveSeriesII}
\end{equation}
If $\rho$ varies on a lengthscale much longer than the characteristic lengthscale of $f_s$, its successive  derivatives will be smaller and smaller.
Assuming a nonzero first moment, we find that 
\begin{equation}
  v(x) \approx -M_1[f_s] \rho_x + {\cal O} (\rho_{xxx}).
\end{equation} 
The successively smaller error terms in (\ref{vLongWaveSeriesII}) correspond to higher-order (positive or negative) diffusion. With the velocity in this form, the governing equation (\ref{eqnOfMotion}) becomes
\begin{equation}
\label{PorousMedium}
\rho_t = \kappa (\rho^2)_{xx},  \quad \kappa = \frac{1}{2} M_1[f_s].
\end{equation}

For $\kappa > 0$, this is the well-known porous-medium equation.  For certain initial conditions, a class of similarity solutions known as Barenblatt solutions are given by
\begin{equation}
  \rho(x,t) = \frac{3^{1/3} M^{2/3}}{4[\kappa(t-t_0)]^{1/3}}
   \left[ 1- \left( \frac{x-x_0}{[9M\kappa(t-t_0)]^{1/3}} \right)^2 \right]_{+},
  \label{barenblattSolution}
\end{equation}
where we use the notation $[u]_+ = \max \{0, u\}$, and where $M$ is the mass (\emph{cf.} Appendix~A) , $x_0$ is its center of mass and $t_0$ is a parameter depending on the initial condition ~\cite{wb1998,zb1958}.  Additionally, all initial conditions for (\ref{PorousMedium}) will approach this particular class of solutions asymptotically as $t \to \infty$. The solutions spread and grow wider without bound. For $\kappa < 0$, (\ref{PorousMedium}) describes backwards diffusion; mathematically, this problem is ill-posed. 

The case $\kappa > 0$ is asymptotically consistent; that is, long-wave states in (\ref{eqnOfMotion}) will spread and therefore remain long-wave when $\kappa > 0$.  Eq. (\ref{PorousMedium}) will become an increasingly valid approximation of (\ref{eqnOfMotion}), justifying \emph{a posteriori} the longwave expansion. However, when $\kappa < 0$, long-wave initial conditions will contract until they can no longer be considered long-wave, at which point the approximations used above become invalid. Finally, we note that if $\kappa = 0$ (that is, if the first moment of $f_s$ vanishes), the above analysis does not hold, and we must retain higher-order terms in (\ref{vLongWaveSeries}) in order to predict asymptotic behavior.

For comparison purposes, we note that the root-mean-square (RMS) width of the Barenblatt solution (\ref{barenblattSolution}) can be computed by first computing the second moment around the center of mass,
\begin{equation}
Q \equiv\int_{-\infty}^{\infty} (x-x_0)^2 \rho(x,t) \, dy = \frac {3^{4/3}}{5} M^{5/3} \kappa^{2/3} (t+t_0)^{2/3}
\end{equation}
which yields
\begin{equation}
RMS = \sqrt{Q/M} = \frac {3^{2/3}}{\sqrt{5}} M^{1/3} \kappa^{1/3} (t+t_0)^{1/3}.
\label{RMS}
\end{equation}
We verify this prediction in Section \ref{spreading} below.

\subsection{Short-wave behavior}
\label{sec:shortwave}

We now consider the evolution of initial conditions that are narrow and sharply-varying. We show that the cumulative density behaves, at least for a short time, according to Burgers' equation \cite{bv2006}. We first define the cumulative mass function:
\begin{equation}
  \psi(x,t) = \int_{-\infty}^{x} \rho(z,t) \, dz.
\end{equation}
Note that since $\rho \geq 0$, $\psi(x)$ increases monotonically from a value of $0$ to $M$. We now use (\ref{eqnOfMotion}) to write
\begin{subequations}
\begin{eqnarray}
  \psi_t(x,t) &=& \int_{-\infty}^{x} \rho_t(z,t) \, dz \\
    &=& - \int_{-\infty}^{x} (\rho(z,t) v(z,t))_z \, dz \\
    &=& - \rho(x,t) v(x,t)  \\
    &=& - \psi_x(x,t) v(x,t).
\end{eqnarray}
\end{subequations}
That is, the cumulative mass function obeys
\begin{equation}
\label{eq:preburgers}
  \psi_t + v \psi_x = 0.
\end{equation}

To proceed, recall from Section \ref{sec:introduction} our assumption about the social interaction force $f_s$, namely that  $f_s$ is continuous  and piecewise differentiable everywhere except for a jump discontinuity of size $2\beta$ at the origin. Following \cite{bv2006}, we write
\begin{equation}
\label{eq:divide}
f_s(x) =  2\beta H(x) + g(x)
\end{equation}
where $\beta \neq 0$, $H$ is the Heaviside function, and $g(x)$ is continuous and differentiable. Substituting (\ref{eq:divide}) into (\ref{eq:preburgers}) and using the fact that $H*\rho = \psi$ yields
\begin{equation}
\label{eq:preburgers2}
\psi_t + \left(2\beta \psi + g*\rho\right)\psi_x = 0.
\end{equation}
For convenience, and without loss of generality, let the (conserved) center of mass of $\rho$ be at $x=0$. Assume that $\rho$ is initially short-wave, so that $\widehat{\rho} \approx M$ in Fourier space. In this case, the term $(g*\rho)\psi_x \approx Mg\psi_x$. Since $\psi_x \approx \rho$ is short-wave, we may further approximate this term as $Mg(0)\psi_x$. Using the fact that $g(0)=\beta$ from (\ref{eq:divide}) and substituting into (\ref{eq:preburgers2}), we have (approximately) that
\begin{equation}
\psi_t + \left(2\beta \psi + M\beta\right)\psi_x = 0
\end{equation}
for short-wave solutions. This is Burger's equation with an additional constant velocity term. This term may be eliminated by a simple change of variables, for instance letting $\psi \rightarrow \psi - M/2$ to obtain
\begin{equation}
\label{eq:Burgers}
 \psi_t + 2\beta\psi\psi_x = 0, \quad \beta = \lim_{x \downarrow 0} f_s(x).
\end{equation}
We now invoke standard results for Burgers' equation \cite{w1974}. Since $\psi$ is monotonically increasing in $x$, $\psi$ will contract and form a shock when $\beta < 0$ and will spread when $\beta > 0$.  Moreover, because $\psi_x = \rho$, a shock in $\psi$ is manifest as a $\delta$-function in $\rho$, and so we expect that $\rho$ will blow up when $\beta < 0$ and spread when $\beta > 0$. In fact, under mild conditions on $f_s$, \cite{bv2006} shows global existence of solutions for $\beta>0$ and give examples of finite-time shock formation for $\beta<0$.

The case $\beta < 0$ is asymptotically consistent; that is, short-wave initial conditions in (\ref{eqnOfMotion}) will contract and therefore remain short wave when $\beta < 0$. Eq. (\ref{eq:Burgers}) will become an increasingly valid approximation of (\ref{eqnOfMotion}). However, when $\beta  > 0$, short-wave initial conditions will spread until they can no longer be considered narrow, at which point the approximations used above will fail to hold. Finally, we note that if $\beta = 0$, we are in the degenerate case where $f_s$ is continuous at the origin. In this case, the leading order approximation for (\ref{eqnOfMotion}) would involve antiderivatives of the cumulative mass function $\psi$.

\subsection{Predicting qualitative behavior}

From the results in Sections~\ref{sec:longwave} and~\ref{sec:shortwave}, we expect short waves to blow up when $\beta < 0$ and spread when $\beta > 0$.  Similarly, we expect long waves to contract when $\kappa < 0$ and spread when $\kappa > 0$.  When short waves blow up, we expect the short-wave instability to override the long-wave behavior.  Thus, there are three possible cases, one of which has two sub-cases. We summarize these below.
\begin{itemize}
  \item[(A)] When $\beta > 0$ and $\kappa > 0$, both long and short waves expand, leading to spreading solutions of the type shown in Figure \ref{snapshots}(a). The asymptotic dynamics of the density are governed by the porous medium equation (\ref{PorousMedium}). We analyze this case further in Section \ref{spreading}.
  \item[(B)] When $\beta > 0$ and $\kappa < 0$, short waves spread while long waves contract, leading to steady-state solutions of the type shown in Figure \ref{snapshots}(b). We analyze these solutions in depth in \cite{bt2008}.
  \item[(C)] When $\beta < 0$, short waves contract and solutions blow-up regardless of the value of $\kappa$, leading to solutions of the types shown in Figure \ref{snapshots}(cd). The asymptotic dynamics of the cumulative density are governed by Burgers' equation (\ref{eq:Burgers}). We analyze this case further in Section \ref{sec:blowup}.
  \end{itemize}
Note that, plausibly, there could be other asymptotic behaviors that we have not discovered. However, the value of $\beta$ at the origin governs whether organisms are repulsive or attractive at short distances and clearly governs the formation of clumps.  If $\beta>0$,  $\rho$ must spread to at least a width where the long-range attractive forces play a significant role. Our interpretation of $\kappa$ is as a measure of whether long-range attraction can balance short range repulsion (the case when $\kappa<0$) or if the short-range repulsion always dominates (when $\kappa>0$). While more exotic behaviors might be possible with more exotic choices of $f_s$ -- say with multiple bands of attraction and repulsion -- the classification above captures the behaviors observed with the simple, biological choices of $f_s$ considered in this paper.

As an example we consider the class of social forces (\ref{reproducing}), for which $\kappa = 1 - FL^2$ and $\beta = 1 - F$. (Note that the regime $F>1$, $L>1$ corresponds to purely attractive social forces, and the regime $F<1$, $L<1$ corresponds to purely repulsive social forces.) We expect to see blow-up when $F > 1$, spreading when $F < 1/L^2$ and $F < 1$, and steady-state solutions when $1 > F > 1/L^2$. These predictions are indicated in Figure \ref{regimes} which shows $F$-$L$ parameter space. The blow-up boundary $\beta = 1 - F = 0$ is the solid line and the spreading/steady-state boundary $\kappa = 1 - FL^2 = 0$ is the solid curve. The symbols in Figure \ref{regimes} summarize the results of numerical simulations conducted for the particular case when $f_s$ is the Morse function (\ref{morse}). The theoretical curves divide the numerical results as expected. 
Our Figure {\ref{regimes} is similar to ``phase diagrams'' showing the linear stability and statistical mechanical H-stability of other swarming models with Morse-type social forces in \cite{cdmbc2007,dcbc2006}.

In the blow-up regime, we in fact observe two different types of blow-up in the numerical simulations; the boundary between these is indicated as the broken vertical line. The particular form of the blow-up depends on the long-range character of the social force $f_s$. For the class of kernels (\ref{reproducing}), when $L > 1$, $f_s(x) \to 0^-$ as $x \to \infty$ so social forces are attractive at long distances. In this case, the entire mass of the system eventually collapses into a single $\delta$-function. In the other case $L < 1$, $f_s(x) \to 0^+$ as $x \to \infty$ and social forces are repulsive at long distances (a behavior which does not have an immediate biological interpretation). Blow-up still occurs due to the contraction of short waves. However, the long-range repulsion means that the solution does not aggregate into one clump. Instead, multiple $\delta$-functions form which repel each other and move apart.

\begin{figure}[ht] 
\centerline{\epsfig{file=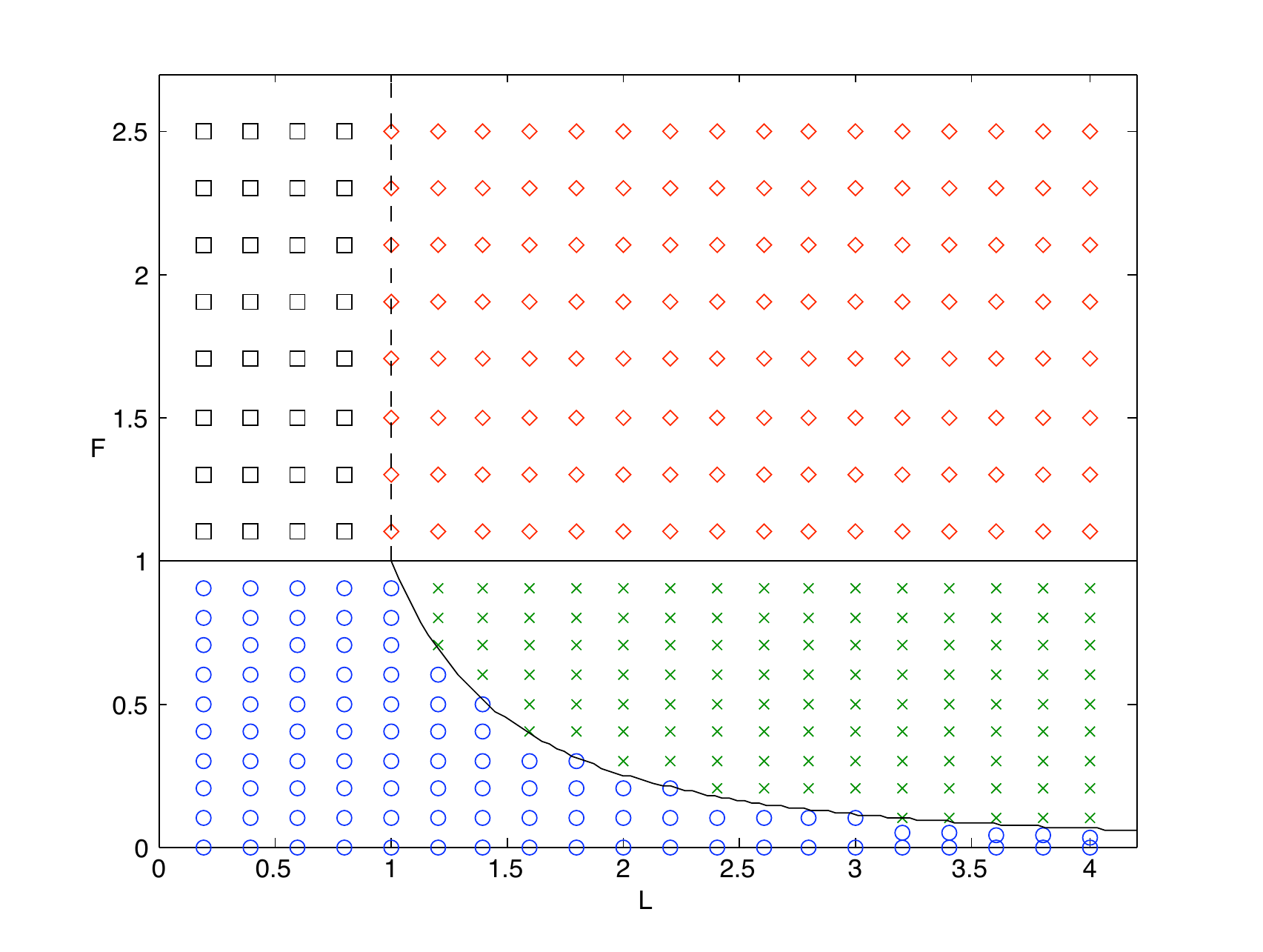, scale=0.8}}
\caption{Different dynamical regimes of the governing equation (\ref{eqnOfMotion}) in $F$-$L$ parameter space for social forces of the form (\ref{reproducing}). $F$ is the relative strength of attraction to repulsion, and $L$ is the relative length scale.  The solid horizontal line at $F = 1$ indicates $\beta = 0$ in (\ref{eq:Burgers}) and marks the theoretical boundary above which solutions blow up. The solid curve $F = 1/L^2$ indicates $\kappa = 0$ in (\ref{PorousMedium}) and marks the theoretical boundary between spreading and steady state solutions for $F < 1$. Results of numerical simulations using Morse-type social forces (\ref{morse}) are indicated by the symbols: spreading ($\circ$), steady-state (x), and blow-up ($\square$ and $\diamond$). The (partial) line $L = 1$ indicates the critical case separating whether the long-range behavior is attractive or repulsive. In the blow-up regime with long-range attraction ($L > 1$), the solution forms a single $\delta$-function ($\diamond$). With long-range repulsion ($L < 1$), multiple, mutually-repelling $\delta$-functions form ($\square$).}
\label{regimes} 
\end{figure}

To verify our analytical results further, we consider a second example with a social force not of the form of (\ref{reproducing}), namely
\begin{equation}
\label{linearpotential}
f_s(x) = \sgn(x) \cdot
\begin{cases}
-\frac{\widetilde{F}+1}{\widetilde{L}}|x| + \widetilde{F} & 0 < |x| \le \widetilde{L} \\
\frac{1}{1-\widetilde{L}}|x| - \frac{1}{1-\widetilde{L}} & \widetilde{L} < |x| \leq 1\\
0 & |x| > 1
\end{cases}
\end{equation}
where $\widetilde{F} \in \mathbb{R}$ and $\widetilde{L} \in (0,1)$. (Note that the regime $\widetilde{F}<0$ corresponds to purely attractive social forces within the range of sensing.) A schematic picture is shown in Figure \ref{socialforces}(b) for the case $\widetilde{F} > 0$. For $x > 0$, this function consists of the compactly supported, piecewise linear function passing through the points $(0,\widetilde{F})$, $(\widetilde{L},-1)$, and $(1,0)$. For $x < 0$ the function is the odd extension. The parameter $\widetilde{F}$ plays a role somewhat similar to $F$ in (\ref{morse}) in that it determines whether the kernel is attractive or repulsive for short distances. The parameter $\widetilde{L}$ plays a role somewhat similar to $L$ in that it determines a characteristic length scale. Eq. (\ref{linearpotential}) differs from (\ref{morse}) in that the kernel is compactly supported rather than decaying, is linear rather than exponential, and by construction is attractive (negative) at intermediate distances regardless of parameter choices.

For (\ref{linearpotential}), $\kappa = (\widetilde{F}\widetilde{L}^2 - \widetilde{L} - 1)/6$ and $\beta = \widetilde{F}$. We expect to see blow-up when $\widetilde{F} < 0$, spreading when $\widetilde{F} > \widetilde{L}^{-2} + \widetilde{L} ^{-1} > 0$, and steady-state solutions when $0 < \widetilde{F} < \widetilde{L}^{-2} + \widetilde{L} ^{-1}$. These predictions are indicated in Figure \ref{regimes2} which is similar to Figure \ref{regimes} except that now we use a social force given by (\ref{linearpotential}) rather than (\ref{morse}). As before, numerical simulations produce spreading solutions and steady-states, both with jump discontinuities at the edges, as well as solutions that blow up. The theoretical predictions for these different regimes (curves) again divide the numerical results, as expected. For this example, only single $\delta$-function blow-up occurs because $f_s < 0$ at intermediate distances and our initial conditions have sufficiently narrow support. Since $f_s$ is compactly supported, initial conditions that are sufficiently wide (or consist of sufficiently distant, separated groups) would blow-up into multiple $\delta$-functions that are stationary, rather than mutually repelling.

\begin{figure}[ht] 
\centerline{\epsfig{file=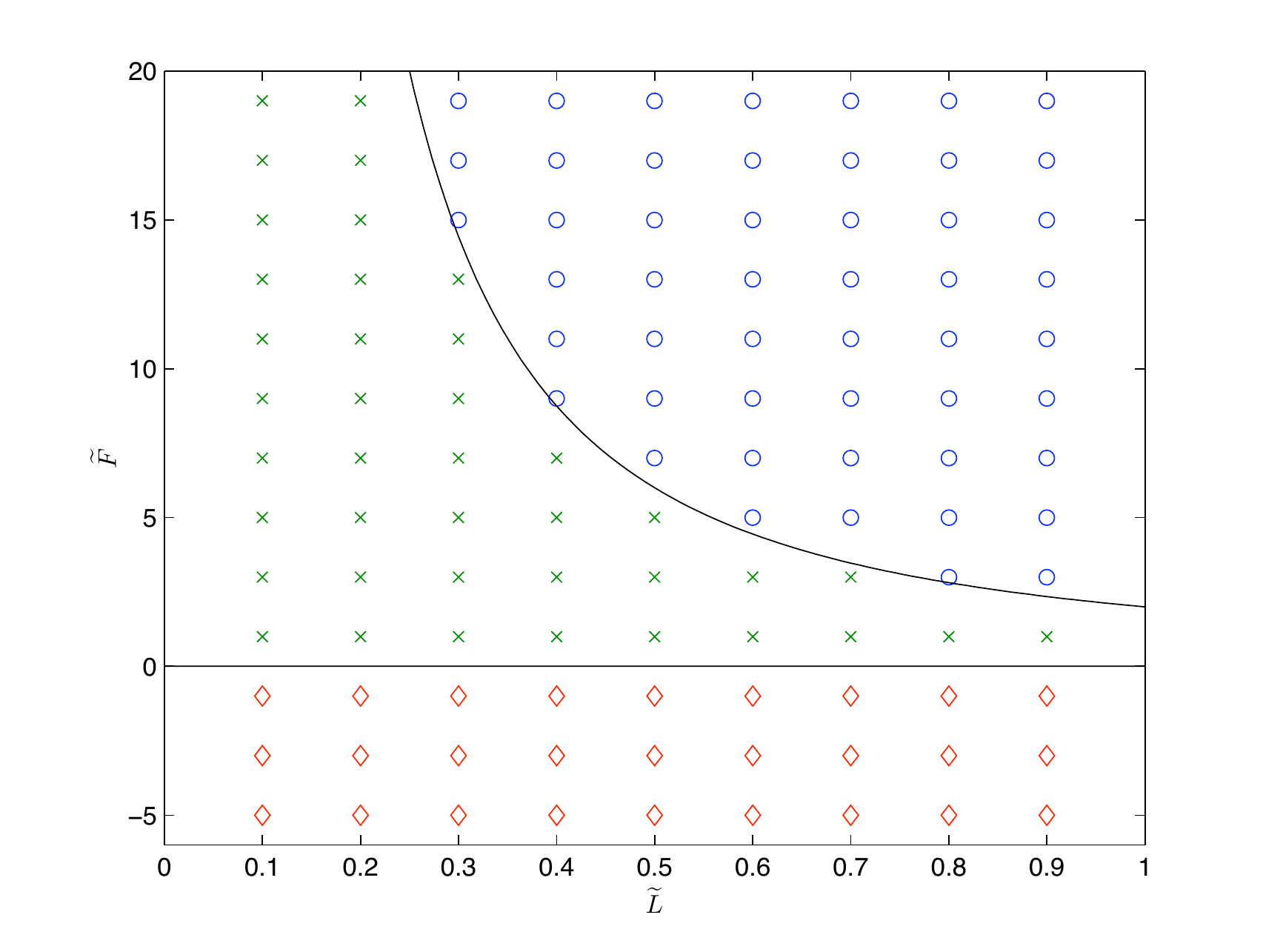, scale=0.8}}
\caption{Different dynamical regimes of the governing equation (\ref{eqnOfMotion}) in $\widetilde{F}$--$\widetilde{L}$ parameter space for the social force (\ref{linearpotential}). The horizontal line at $\widetilde{F} = 0$ indicates $\beta = 0$ in (\ref{eq:Burgers}) and marks the theoretical boundary below which solutions blow up. The solid curve $\widetilde{F} = \widetilde{L}^{-2} + \widetilde{L}^{-1}$ indicates $\kappa = 0$ in (\ref{PorousMedium}) and marks the theoretical boundary between spreading and steady state solutions. Results of numerical simulations  are indicated by the symbols: spreading ($\circ$), steady-state (x), and single-$\delta$-function blow-up ($\diamond$).}
\label{regimes2} 
\end{figure}

\section{Spreading Solutions}
\label{spreading}

When $\kappa > 0$ in (\ref{PorousMedium}) and $\beta > 0$ in (\ref{eq:Burgers}), solutions will spread.  As discussed in Section \ref{sec:predictions}, the density profile grows wider, the long-wave approximation (\ref{PorousMedium}) will become increasingly accurate, and so we expect solutions to approach Barenblatt's solution (\ref{barenblattSolution}). Figures \ref{barenblatt1} and \ref{barenblatt2} confirm this prediction.

Figure \ref{barenblatt1} compares Barenblatt's solution (\ref{barenblattSolution}) to numerical simulations of (\ref{eqnOfMotion}) using the Morse-type social force (\ref{morse}) with $F=0.2$ and $L = 2$. For these parameters, $\kappa = 0.2 > 0$ in (\ref{PorousMedium}). The broken line represents Barenblatt's solution (\ref{barenblattSolution}) for a density profile with unit mass. Under the similarity transformation
\begin{equation}
\label{rescaling}
\widetilde{\rho}(\widetilde{x}) = \frac{1}{\gamma} \rho (\gamma x), \quad \gamma = \max_x \rho(x)
\end{equation}
the spreading Barenblatt profile collapses to a single curve. The solid lines represent snapshots from the numerical simulation of (\ref{eqnOfMotion}). We apply to these numerical snapshots the same rescaling (\ref{rescaling}). We take the initial condition to be a rectangular pulse with unit mass. The direction of increasing time is indicated by the arrow in the figure. As time increases, the numerical profiles, as expected, approach the Barenblatt profile. We explore this approach further in Figure  (\ref{barenblatt2}), which compares the root-mean-square (RMS) width of the solution. From (\ref{RMS}), the RMS width should grow as $t^{1/3}$. As predicted, the RMS width of the numerical solution (circles) approach the theoretical curve (line) on the log-log plot.

\begin{figure}[ht] 
\centerline{\epsfig{file=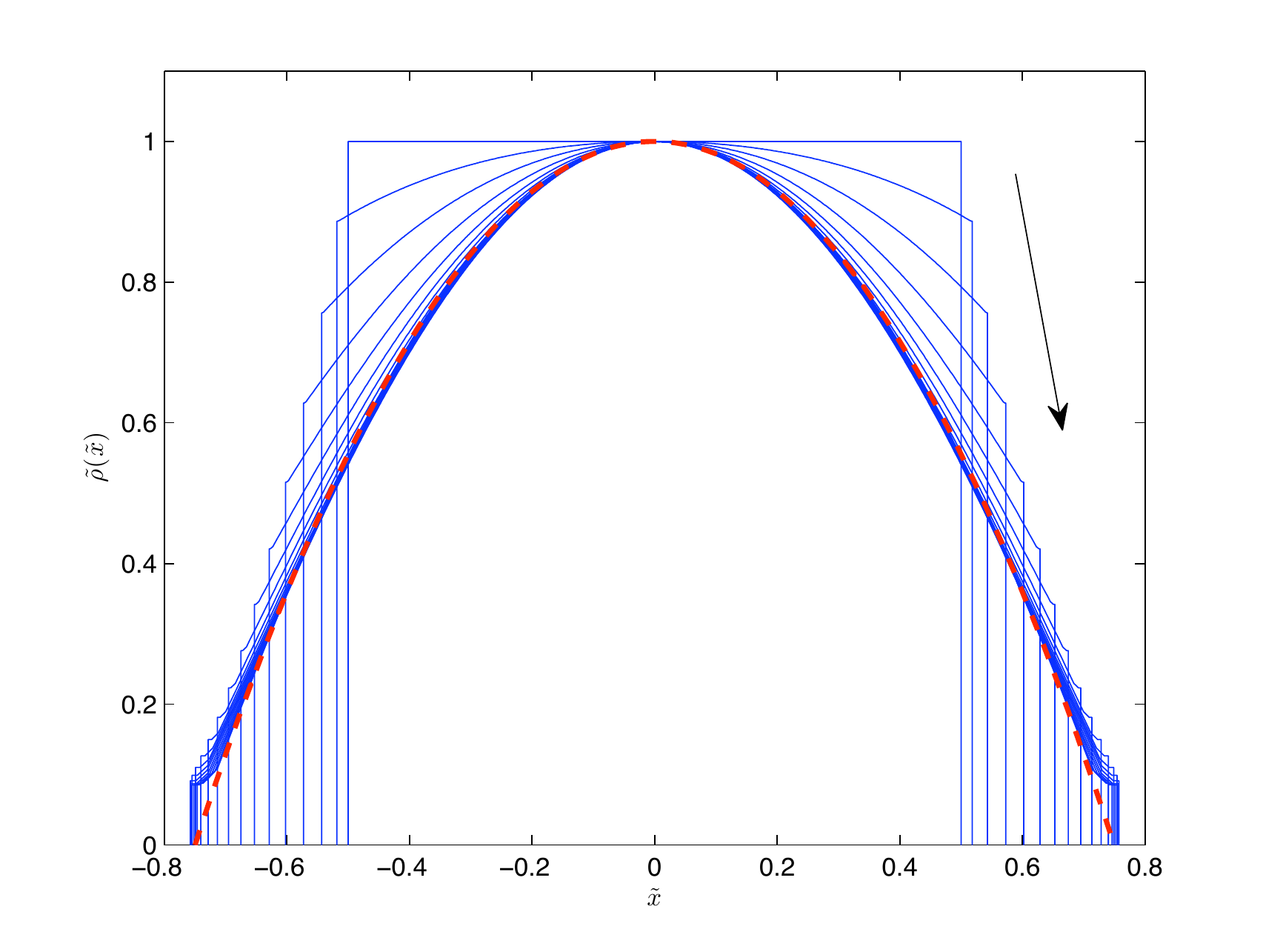, scale=0.8}}
\caption{Spreading solutions to (\ref{eqnOfMotion}) using the Morse-type social interaction (\ref{morse}) with $F=0.2$ and $L = 2$. We choose as an initial condition a rectangular pulse with unit mass. Snapshots of the evolving profile are rescaled according to the similarity transformation (\ref{rescaling}). These evolving solutions are the solid curves and the arrow indicates the time evolution. As predicted, the numerical solutions approach the idealized Barenblatt similarity solution (\ref{barenblattSolution}), which has been similarly rescaled and is shown as the broken curve.}
\label{barenblatt1} 
\end{figure}

\begin{figure}[ht] 
\centerline{\epsfig{file=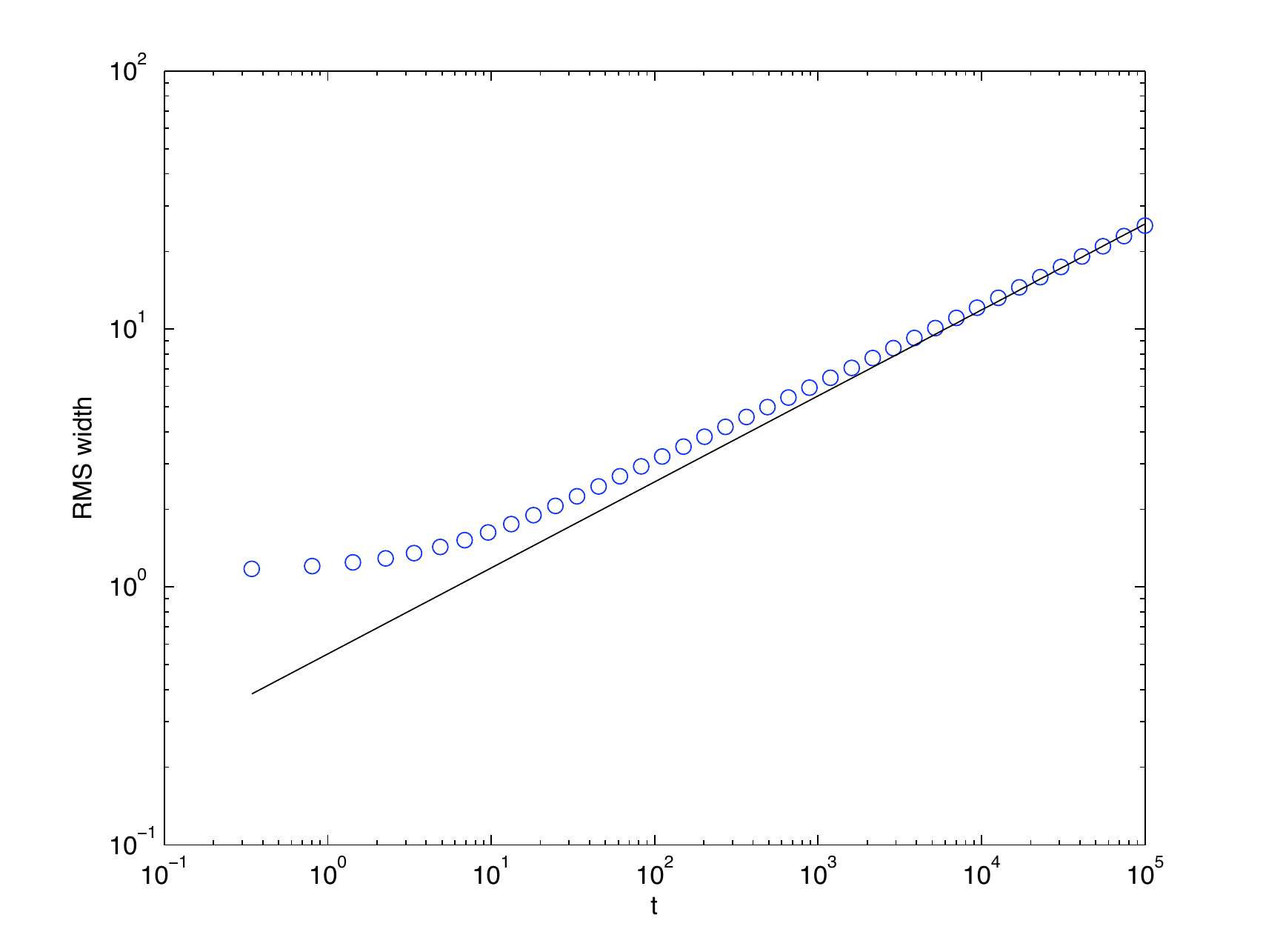, scale=0.8}}
\caption{RMS width of the solutions in Figure \ref{barenblatt1}. The solid line represents Barenblatt's solution and has equation 
$RMS = 3^{2/3}5^{-5/6} (t+t_0)^{1/3} \approx 
0.544 t^{1/3}$ at large times (\ref{RMS}). Circles represent the numerical solutions to (\ref{eqnOfMotion}) which asymptotically approach the Barenblatt spreading rate.} 
\label{barenblatt2} 
\end{figure}

Apart from the shape of the solution, we also wish to study the jump discontinuities at the edge of the spreading swarm. At the left edge of the swarm, we might expect a spreading solution to behave locally like a fixed wave profile traveling to the left (and similarly at the right edge). Therefore, we  seek a traveling-wave solution to (\ref{eqnOfMotion}).  At the left endpoint,  we look for a traveling-wave solution of the form $\rho(x,t) = g(x+ct)$, where $g(z) = 0$ for all $z<0$. Under these assumptions, (\ref{eqnOfMotion}) reduces to
\begin{equation}
  0 = c\pd{g}{z} + \pd{}{z}(vg) = \pd{}{z}[(c+v)g].
\end{equation}
Integrating both sides of this equation,
\begin{equation}
  (c+v)g = 0.
\end{equation}
The constant of integration is zero because the left-hand side vanishes for negative $z$.  Hence, wherever $g$ is nonzero, $-c = v$.  That is,
\begin{equation}
  -c = \int_0^\infty g(\zt) f_s(z - \zt) \, d\zt \quad \textrm{for $z \geq 0$}.
  \label{eqn:TWIntMorse1}
\end{equation}

We proceed with a quantitative analysis for the example case of Morse-type social interactions (\ref{morse}). Writing out  $f_s$ explicitly and taking derivatives with respect to $z$ on both sides (which eventually facilitates transformation of the integral equation into an ODE) yields, after some rearranging,
\begin{equation}
  (F-1)g(z) = \frac{1}{2} \int_0^\infty g(\zt) \left[ \frac{F}{L} e^{-|z-\zt|/L}
         - e^{-|z-\zt|} \right] \, d\zt
  \label{eqn:TWIntMorse2}
\end{equation}
To ensure that the exponential terms are linearly independent, we assume $F \neq 0$ and $L \neq 1$.  Then, to solve this integral equation, we apply the differential operators $\mathcal{L}_1 = \partial_{zz} - 1$ and $\mathcal{L}_2 = L^2 \partial_{zz} - 1$ to both sides. The left-hand side becomes
\begin{subequations}
\begin{eqnarray}
  &&\mathcal{L}_1 \mathcal{L}_2 [(F-1)g(z)] \\
    & = & (F-1) L^2 g''''(z) + (1-F+L^2-FL^2) g''(z) + (F-1) g(z),
\end{eqnarray}
\end{subequations}
and the right-hand side becomes
\begin{subequations}
\begin{eqnarray}
  &&\frac{1}{2} \int_0^\infty g(\zt) \mathcal{L}_1 \mathcal{L}_2 \left[ \frac{F}{L} e^{-|z-\zt|/L} - e^{-|z-\zt|} \right] \, d\zt \\
 &=& \frac{1}{2} \int_0^\infty g(\zt) \cdot (-2) \cdot \left[ (F-L^2) \delta''(z-\zt) + (1-F)\delta(z-\zt) \right] \, d\zt \\
 &=& (L^2-F)g''(z) + (F-1)g(z).
\end{eqnarray}
\end{subequations}

Hence, the integral equation (\ref{eqn:TWIntMorse1}) reduces to the ODE
\begin{equation}
  g''''(z) - \alpha^2 g''(z) = 0
\end{equation}
where
\begin{equation}
\alpha^2 =  \frac{1-FL^2}{L^2(1-F)}.
\end{equation}
Because we are in the spreading regime by assumption, $1-FL^2 = \kappa > 0$ and $1-F = \beta > 0$, so $\alpha^2 > 0$ and thus the coefficient on $g''(z)$ is strictly negative. Integrating the ODE twice yields
\begin{equation}
  g''(z) - \alpha^2 [g(z) - Az - B] = 0
\end{equation} 
which has general solution
\begin{equation}
\label{eq:gensoln}
  g(z) = Az + B + Ce^{-\alpha z} + De^{\alpha z}.
\end{equation}
The traveling wave cannot grow exponentially as $z\to\infty$, as this would imply exponentially growing mass flux  (which is proportional to the product of the speed and the derivative of the profile) as the wave translates to the left, so we choose $D=0$.  To find $A$, $B$, and $C$ we plug (\ref{eq:gensoln}) into (\ref{eqn:TWIntMorse1}) and simplify to obtain
\begin{eqnarray}
  -c &=& A\cdot\left[2(FL^2 - 1) - FL^2 e^{-z/L} + e^{-z}\right] \\
     &&\mbox{} + B\cdot\left[FLe^{-z/L} - e^{-z}\right]  \nonumber \\
     &&\mbox{} + C\cdot\left[\frac{FL}{1-\alpha L}e^{-z/L} - \frac{1}{1-\alpha}e^{-z}\right]. \nonumber
\end{eqnarray}

Since $\{1,e^{-z/L},e^{-z}\}$ are linearly independent, we can match like terms and solve the resulting three algebraic equations for $A$, $B$, and $C$ to obtain
\begin{subequations}
\label{eq:abc}
\begin{eqnarray}
  A &=& cA_0, \quad A_0 = \frac{1}{2(1-FL^2)} \\
  B &=& cB_0, \quad B_0 = \frac{1}{2(1-FL^2)} \left( L + 1 - \frac{1}{\alpha} \right) \\
  C &=& cC_0, \quad C_0 = \frac{1}{2(1-FL^2)} (\alpha L - 1) \left( 1 - \frac{1}{\alpha} \right)
\end{eqnarray}
\end{subequations}
which determines a traveling-wave solution for each wave speed $c$.

Figure \ref{TWS2} shows an example of the traveling left edge of the spreading swarm studied in Figures \ref{barenblatt1} and \ref{barenblatt2}. We plot three snapshots of the numerically spreading solution and superpose the analytical solution given by (\ref{eq:gensoln}) and (\ref{eq:abc}). The two are in good agreement close to the edge of the support where the traveling wave calculation above is expected to be valid. 

\begin{figure}[ht] 
\centerline{\epsfig{file=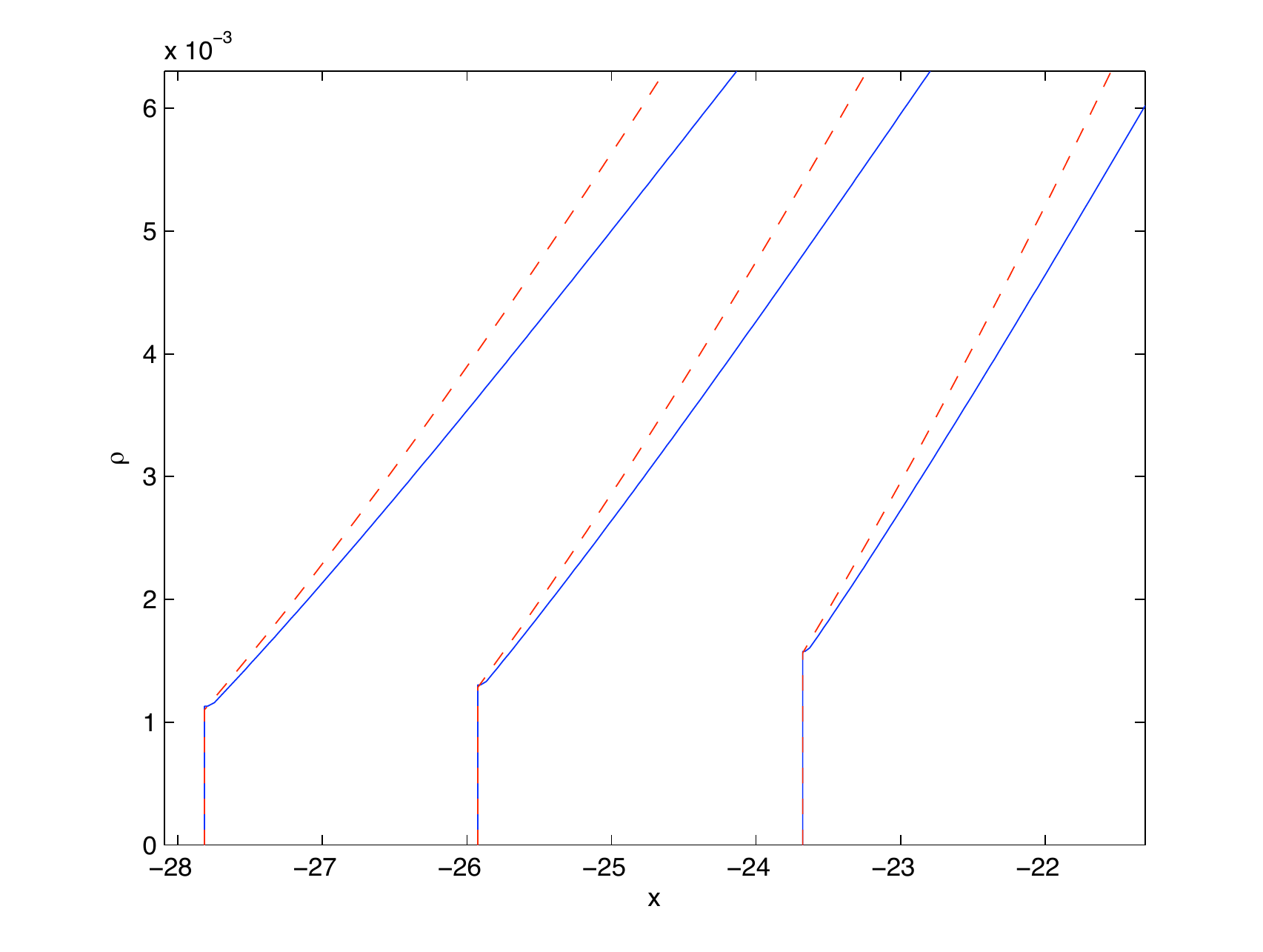, scale=0.8}}
\caption{Evolving left edge of the spreading solution studied in Figures \ref{barenblatt1} and \ref{barenblatt2}. We superpose the analytical solution (dashed curves) given by (\ref{eq:gensoln}) and (\ref{eq:abc}) on the numerical solution (solid curves). The two are in good agreement close to the edge of the support where the traveling wave calculation in Section \ref{spreading} is expected to be valid.} 
\label{TWS2} 
\end{figure}

To verify (\ref{eq:gensoln}) and (\ref{eq:abc}) further, we predict the relationship between the instantaneous speed of a traveling front, the size of the jump at the edge, and the slope of the density at the edge. In particular, the size of the jump is
\begin{equation}
\label{eq:twpred1}
  g(0) = c(B_0 + C_0),
\end{equation}
and the slope at the edge is
\begin{equation}
\label{eq:twpred2}
  g'(0) = A - \alpha C e^{-\alpha z} \big|_{z=0} = c(A_0-\alpha C_0).
\end{equation}
A similar analysis holds at the opposite edge of the swarm. We expect these relations to hold only for large $t$ since the solution must be sufficiently wide and slowly varying near the endpoints for it to locally approximate a traveling wave. For several values of $F$ and $L$, we tested these predictions by tracking the speed, jump in density, and slope at the endpoints over time.  Figure \ref{TWS} shows an example that confirms the traveling wave predictions. This example corresponds to the same spreading profile studied in Figures \ref{barenblatt1} and \ref{barenblatt2}. Denote the location of the left edge of the swarm by $x_e$. We plot three ratios involving quantities computed at the edge, namely
\begin{equation}
\label{eq:ratios}
\frac {\rho(x_e)} {c(B_0+C_0)}, \quad \frac{\rho_x(x_e)} {c(A_0-\alpha C_0)}, \quad \frac {\rho(x_e)(A_0-\alpha C_0)} {\rho_x(x_e)(B_0+C_0)}
\end{equation}
where we take as the values of $\rho(x_e)$ and $\rho_x(x_e)$ their limit approaching from the inside of the support. Each of the three quantities in (\ref{eq:ratios}) approaches unity as $t \to \infty$ as predicted by (\ref{eq:twpred1}) and (\ref{eq:twpred2}).

\begin{figure}[ht] 
\centerline{\epsfig{file=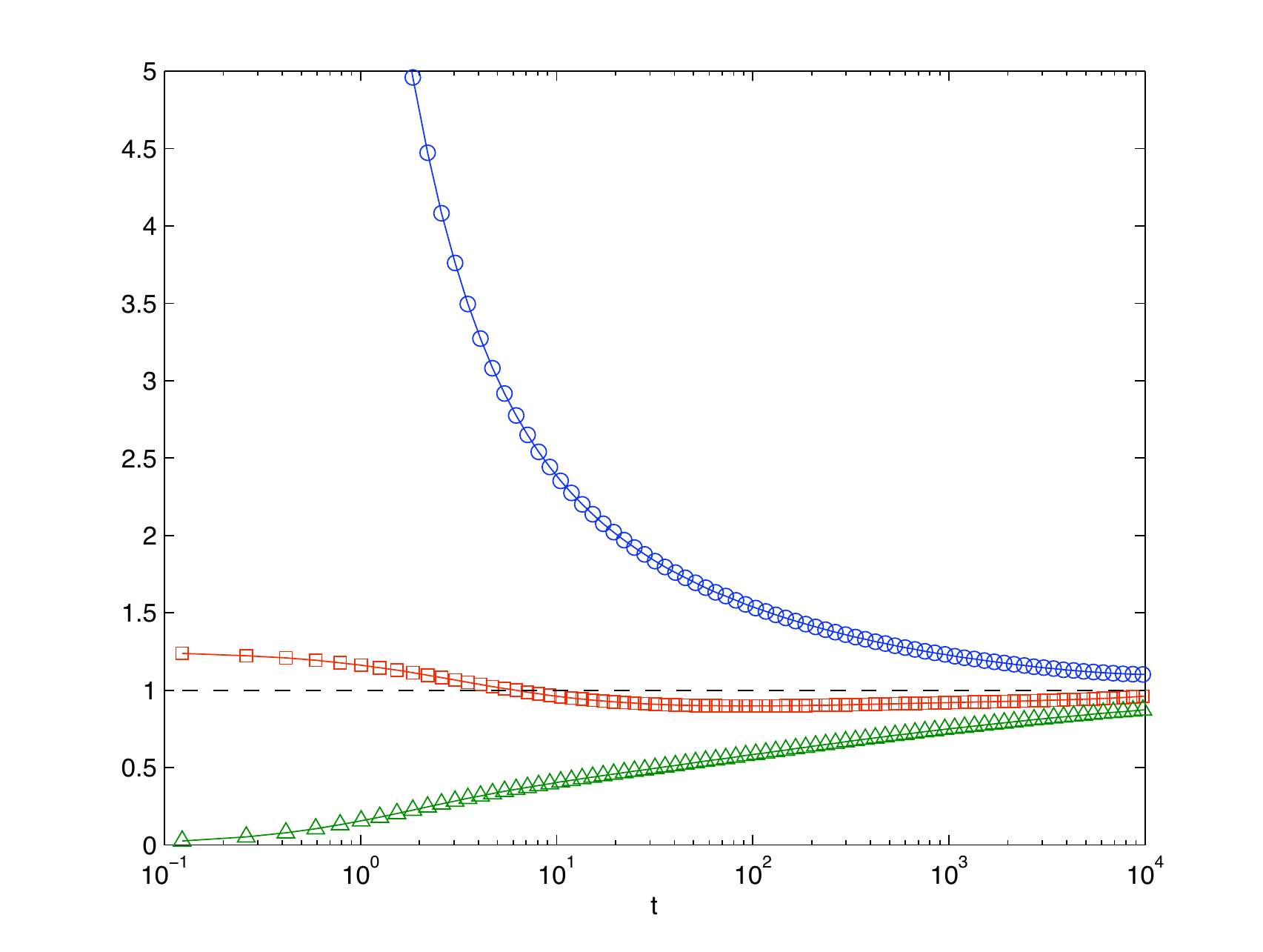, scale=0.8}}
\caption{Edge behavior of a spreading solution. This example corresponds to the profile studied in Figures \ref{barenblatt1} and \ref{barenblatt2}. The squares, triangles, and circles indicate, respectively, the three ratios in (\ref{eq:ratios}). These quantities all approach unity (horizontal dotted line) for large $t$, which indicates that the asymptotic dynamics of the endpoints obey the results of the traveling wave analysis. Specifically, as predicted by (\ref{eq:twpred1}) and (\ref{eq:twpred2}), the jump in density and the slope of the density profile at the edge are both proportional (via known constants) to the speed at which the edge moves.} 
\label{TWS} 
\end{figure}

\section{Contracting solutions}
\label{sec:blowup}

We now consider the case when $\beta < 0$ and solutions blow up due to short-wave contraction. Biologically, this means that at short distances, organisms are attracted to each other leading to clumping. As discussed in Section \ref{sec:predictions}, the density profile $\rho$ blows up by forming one or more $\delta$-functions, or equivalently by the cumulative density $\psi$ forming shocks. The space-time plot Figure \ref{shock} shows an example. Lines represent contours of $\psi$ in the simulation of (\ref{eqnOfMotion}). The value is coded by shading, indicating the characteristics of the hyperbolic problem. As we expect, the characteristics intersect and form a shock after sufficient time, corresponding to blow-up of $\rho$. The authors of \cite{bl2007} study (\ref{eqnOfMotion}) for the case when $f_s \leq 0 $ and rigorously show blow-up in finite time. The blow-up profiles for certain other instances of $f_s$ are studied in \cite{bv2006}, which also shows finite-time blow-up for $\beta < 0$. In Section \ref{sec:predictions} we showed blow-up for $\beta < 0$, regardless of initial conditions, for any $f_s$ satisfying our prior assumptions 

\begin{figure}[ht] 
\centerline{\epsfig{file=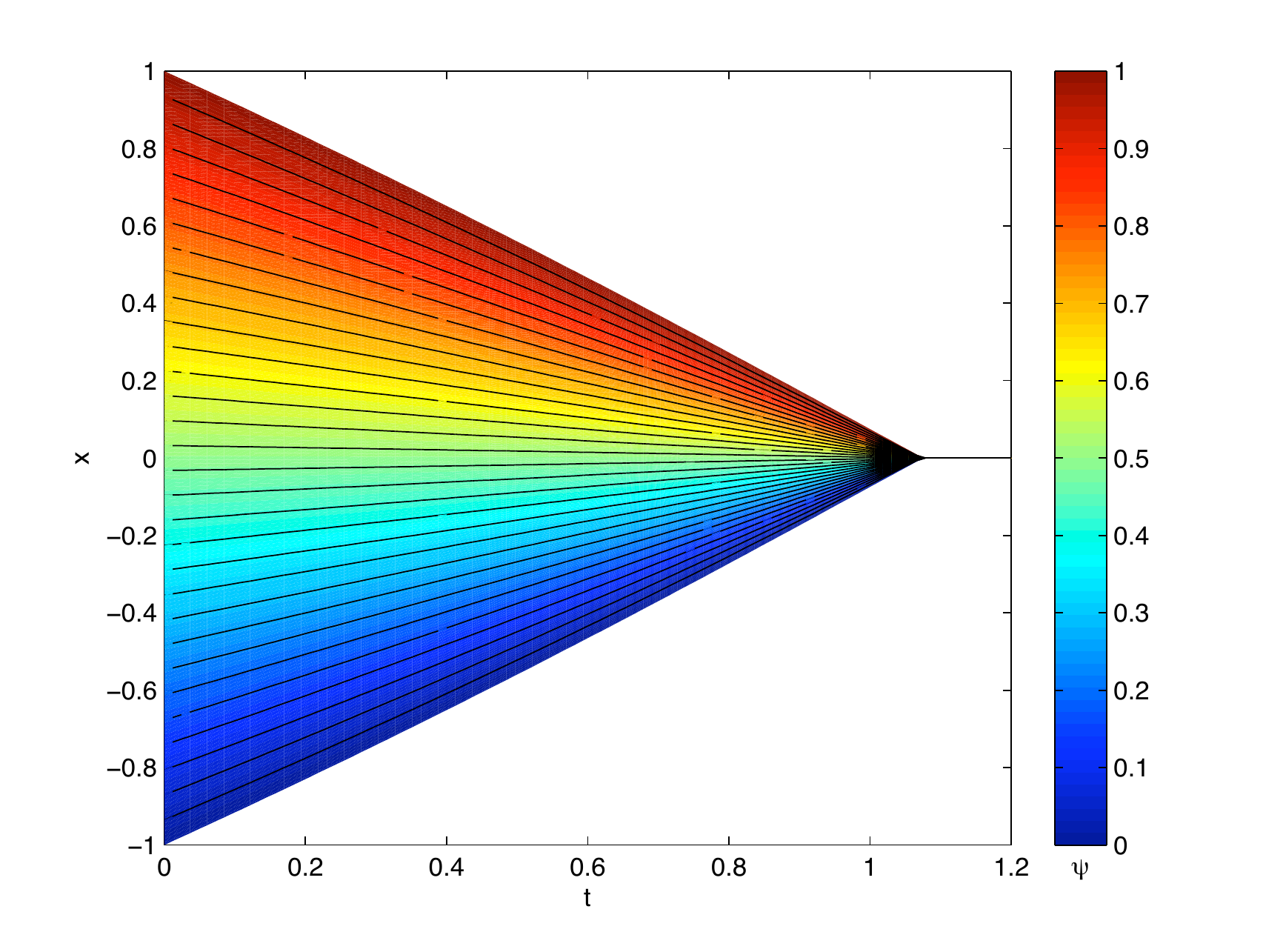, scale=0.8}}
\caption{Space-time plot of the cumulative mass $\psi$ showing blow-up of the density $\rho$ under the dynamics of (\ref{eqnOfMotion}). We use the Morse function (\ref{morse}) with attractive strength $F = 2$ and attractive length scale $2$, for which $\beta = -1$ in (\ref{eq:Burgers}). Contours of $\psi$ appear as lines, and the value is coded by shading, indicated the characteristics of the hyperbolic problem. As we expect, the characteristics intersect and form a shock, which means that $\rho$ blows-up by forming a $\delta$-function as predicted.}
\label{shock} 
\end{figure}

When the initial condition is a single, sufficiently narrow pulse, we can approximately predict when the solution will form a $\delta$-function.  Let $a(t)$ and $b(t)$ denote the position at time $t$ of the left and right edges, respectively.  Then, for $a(t) < z < b(t)$ and $b(t)-a(t)$ sufficiently small, note that
\begin{eqnarray}
  f_s(a(t)-z) &\geq& \min_{a(t)-b(t)<r<0} f_s(r) \\
  &=& \max_{0<r<b(t)-a(t)} f_s(r) \\
  &=& \min_{0<r<b(t)-a(t)} |f_s(r)|.
\end{eqnarray}
Here, we know $f_s(b-a) < 0$ because $b-a$ is small and positive, and  because $\beta = \lim_{x \downarrow 0} f_s(x) < 0$, with $f_s$ continuous except at the origin (\emph{cf.} Section \ref{sec:predictions}). For convenience, define
\begin{equation}
\label{eq:qt}
  q(t) = \min_{0<r<b(t)-a(t)} |f_s(r)|.
\end{equation}
We find a bound for the velocity at the left endpoint:
\begin{eqnarray}
  a'(t) &=& v(a(t), t) \\
  &=& \int_{a(t)}^{b(t)} \rho(z,t) \, f_s(a(t)-z) \, dz \\
  &\geq& q(t) \int_{a(t)}^{b(t)} \rho(z,t) \, dz \\
  &=& Mq(t) \\
  &>& 0.
\end{eqnarray}
A similar argument shows that the velocity at the right endpoint satisfies
\begin{equation}
b'(t) \leq -Mq(t) < 0.
\end{equation}
Thus, the endpoints approach each other.  From (\ref{eq:qt}) it follows that $q(t)$ must be non-decreasing, and consequently the endpoints must be accelerating towards each other, or at least moving towards each other at a constant velocity.

Let $t^*$ denote the time at which all the mass of the system enters a single $\delta$-function. We can find an upper bound for $t^*$ by noting
\begin{equation}
  b'(t) \leq b'(0) \leq -Mq(0), \quad a'(t) \geq a'(0) \geq Mq(0).
\end{equation}
Since the endpoints are initially separated by a distance $b(0) - a(0)$ and each is moving towards the other at a minimum speed $Mq(0)$, this gives the bound
\begin{equation}
  t^* \leq \frac{b(0) - a(0)}{2Mq(0)}.
  \label{blowupBound}
\end{equation}
Furthermore, just before the solution forms a $\delta$-function, by similar argumentation, the velocities of the endpoints will be $M|\beta|$ at the left endpoint, and $-M|\beta|$ at the right endpoint.  If $\beta = 0$ but attraction dominates at small distances, a careful analysis of $q(t)$ suggests that either finite-time or infinite-time blow-up can occur.

We have studied blow-up in numerical simulations of (\ref{eqnOfMotion}); results appear in Figure \ref{blowup1}. We use the Morse-type social force (\ref{morse}) with $L=2$. Blow-up time is shown as a function of $F$ for two different initial conditions, namely a rectangular pulse of width $0.1$ (circles) and one of width $0.2$ (squares). Both sets of data closely match the analytical upper bound, indicated as solid and broken curves respectively.

\begin{figure}[ht] 
\centerline{\epsfig{file=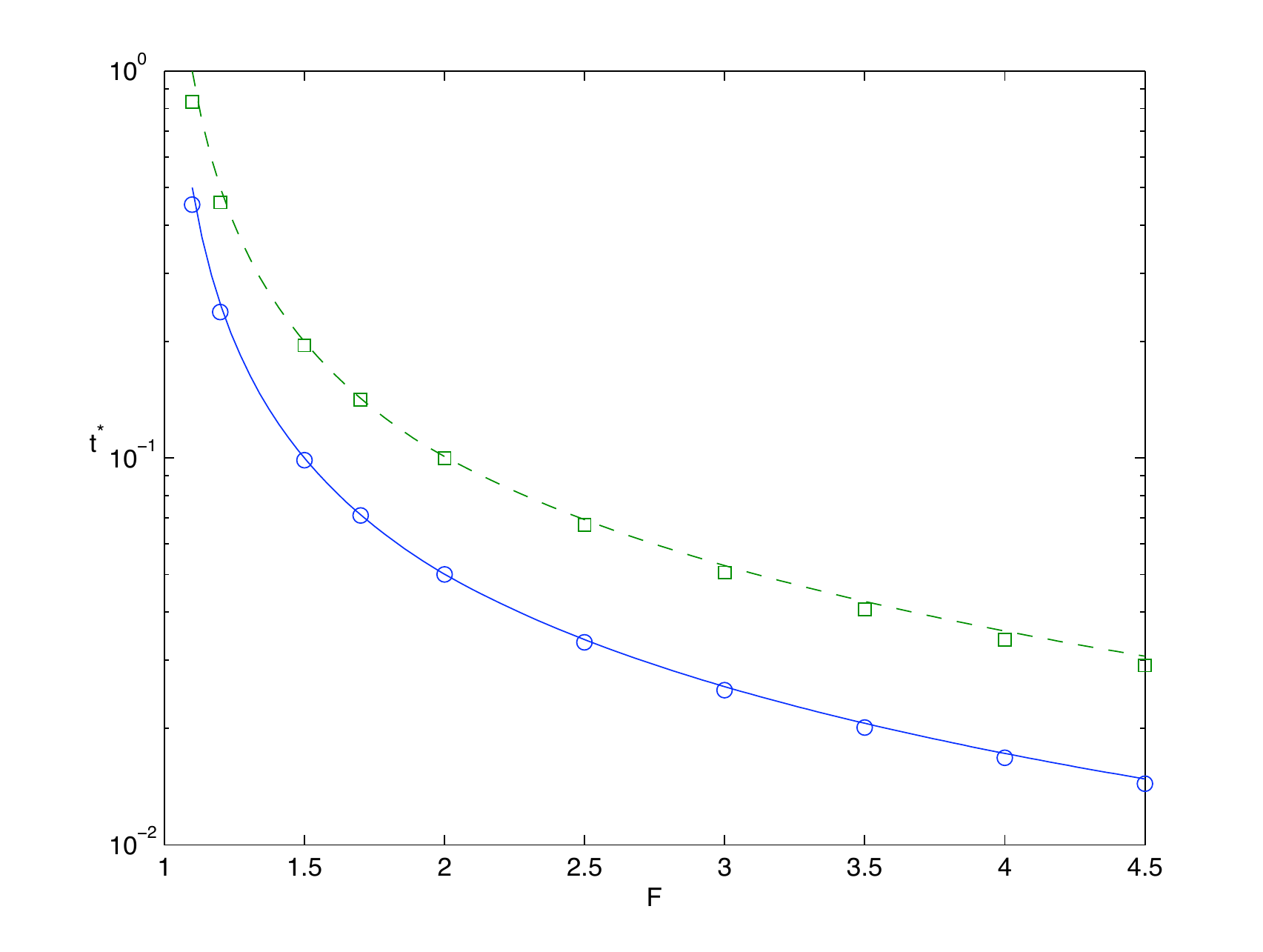, scale=0.8}}
\caption{Blow-up times $t^*$ for (\ref{eqnOfMotion}) with Morse-type interactions (\ref{morse}). We set $L=2$ and vary $F$. Data correspond to two sets of initial conditions, namely a rectangular pulse of width $0.1$ (circles) and one of width $0.2$ (squares). The actual blow-up times are well-approximated by the analytical upper bound (\ref{blowupBound}), indicated as the solid and broken curves.}
\label{blowup1} 
\end{figure}

\section{Conclusions}
\label{sec:conclusions}

In this paper we have studied the swarming-type equation (\ref{eqnOfMotion}) with the goal of predicting how the asymptotic dynamics depend on the social interaction force $f_s$. We consider the class of social interactions that are antisymmetric (describing isotropic interactions), have finite first moments, and are continuous except at the origin. From a long-wave and a short-wave analysis, we showed that two parameters computed directly from $f_s$ determine the asymptotic dynamics, namely the first moment $2 \kappa$ and the limit approaching the origin from the right $\beta$. When $\beta > 0$ and $\kappa > 0$, long and short waves both expand and as $t \to \infty$, the dynamics approach those of the porous medium equation. The shape of the profile and its spreading rate approach those of Barenblatt's well-known  solution. For the case of Morse-type interactions, we calculated a quantitative relationship between the edge speed, edge density, and edge slope.  An interesting question for analysts is to determine conditions on $f_s$ such that this self-similar solution is a global attractor. It is clear that $\beta > 0$ and $\kappa > 0$ are necessary, and we conjecture that our restriction that $f_s$ has at most a single zero crossing (for $x>0$) is sufficient.

In the case $\beta > 0$, $\kappa < 0$, long waves contract and short waves expand. In this case, the system asymptotically reaches a steady state whose shape is highly nontrivial. In~\cite{bt2008}, we analyze these solutions in detail both for the original governing equation (\ref{eqnOfMotion}) as well as for the case when (\ref{velocityEquation}) contains an additional term describing exogenous forces acting on the population (for instance, the effects of gravity, light, or a nutrient field).

When $\beta < 0$, short waves contract. Regardless of the long-wave behavior, solutions blow up as $t \to \infty$. In this limit, the dynamics of the cumulative density obey Burger's equation and form shocks; hence, the density forms one or more $\delta$-functions.

There are several clear directions for future work. First, with the definition of the social force $f_s$ correspondingly modified, many results from Section \ref{sec:predictions} could be extended to the case of higher dimensions. In particular, the predictions of the boundaries of the different dynamical regimes would be of interest. Second, though we have studied in \cite{bt2008} the effect of exogenous forces on steady-state solutions, we have not studied their effect on the spreading and contracting solutions that are the focus of the present paper, nor their potential effect on shifting the dynamical regime boundaries. Other extensions would include the addition of alignment forces and loosening of the restriction of omnidirectional communication, both of which would require an appropriate reformulation of the governing equations of motion. Such a study might shed light on the role parameter choices play in the models investigated in \cite{edl2007,edll2006}.

We hope that our present study will guide mathematicians, biologists and engineers who wish to construct swarming models with particular behaviors that either mimic those observed in nature or are desirable qualities for control of robotic or virtual agents. Our results suggest that although many functional forms can be imagined for the social forces, only a few types of qualitative behavior manifest for this class of model.  From another perspective, selecting a particular functional form to model social interactions is less important than choosing the parameters in that model to manifest a desired behavior. Finally, this study suggests that there are a number of characteristics inherent to kinematic models, namely spreading parabolic profiles, small-scale clumps, and groups with jump discontinuities at the edge, which may be used to diagnose when this class of models is appropriate.

\section*{Acknowledgments}

CMT and AJL acknowledge support from the NSF through grant DMS-0740484. AJL and AJB are grateful to Harvey Mudd College for computation support. Portions of this research were conducted as part of AJL's senior thesis at Harvey Mudd College. We wish to thank both referees for their careful and perceptive comments which greatly improved this paper.

\appendix

\section{Basic properties of the continuum equation}
\label{sec:moments}

We demonstrate that the governing equation (\ref{eqnOfMotion}) conserves mass and center of mass. We take the social force $f_s$ in (\ref{velocityEquation}) to be odd, which describes the case of isotropic social interactions. That is, organisms in disparate locations have a social effect on each other that is equal in magnitude and opposite in direction.

Conservation of mass follows from the fact that (\ref{eqnOfMotion}) is formulated as a conservation law.  To see this explicitly, define the mass of the system at time $t$ as
\begin{equation}
 M(t) = \intR \rho(x,t) \, dx,
\end{equation}
and so
\begin{equation}
  \frac{dM}{dt} = \intR \rho_t \, dx = -\intR (\rho v)_x \, dx = -\big[ \rho v \big]_{x=-\infty}^{x=+\infty} = 0,
\end{equation}
assuming the density decays to zero as $x \to \pm \infty$.  We denote the mass of the system at any time as $M=M(0)$.

Intuition suggests that the center of the mass of the system should also remain fixed because the antisymmetry of the social interaction force $f_s$. To verify this, we consider the center of mass at time $t$:
\begin{equation}
  \bar{x}(t) = \frac{1}{M} \intR x \rho(x,t) \, dx.
\end{equation}
Then,
\begin{eqnarray}
  \frac{d\bar{x}}{dt} 
    &=& \frac{1}{M} \intR x \rho_t \, dx \\
    &=& - \frac{1}{M}\intR x (\rho v)_x \, dx  \\
    &=& - \frac{1}{M}\big[ x \rho v \big]_{x=-\infty}^{x=+\infty}
        + \frac{1}{M} \intR \rho v \, dx.
\end{eqnarray}
Assuming the density vanishes at $\pm \infty$,
\begin{eqnarray}
  \frac{d\bar{x}}{dt}
    &=& \frac{1}{M}\intR \rho v \, dx \\
    &=& \frac{1}{M}\intR \rho(x,t) \intR \rho(y, t) f_s(x - y) \, dy \, dx \\
    &=& \frac{1}{M}\intR \intR \rho(x,t) \rho(y,t) f_s(x-y) \, dy \, dx.
\end{eqnarray}
Relabeling the variables of integration and invoking the antisymmetry of the social interaction force,
\begin{eqnarray}
  \frac{d\bar{x}}{dt}
    &=& \frac{1}{M}\intR \intR \rho(y,t) \rho(x,t) f_s(y-x) \, dx \, dy \\
    &=& -\frac{1}{M}\intR \intR \rho(x,t) \rho(y,t) f_s(x-y) \, dy \, dx.
\end{eqnarray}
Hence, $d\bar{x}/dt = -d\bar{x}/dt$, and thus $d\bar{x}/dt = 0$.  That is, the center of mass is stationary.

\section{Numerical method}
\label{sec:numericalmethod}

Our numerical solution of (\ref{eqnOfMotion}) hinges on a correspondence with a discrete model that approximates it. Ref. \cite{bv2005} shows that a discrete model of the type we will derive converges to the continuous model under fairly weak assumptions.

Our correspondence works as follows. Consider a continuous distribution $\rho_c(x,t)$ with total mass $M$. For ease of notation, we suppress time dependence for the remainder of this paragraph. Define the cumulative density function
\begin{equation}
\psi_c(x) = \int_{x_0}^x \rho_c(s)\,ds
\end{equation}
where the dummy coordinate $x_0$ is taken to the left of the support of $\rho_c$. We seek a discrete approximation of $N$ $\delta$-function point-masses each of mass $m=M/N$. That is,
\begin{equation}
\label{eq:rhod}
\rho_d(x) = \sum_{i=1}^{N} m \delta(x-x_i).
\end{equation}
The associated cumulative density function $\psi_d$ is
\begin{equation}
\psi_d(x) =
\begin{cases}
0 & x < x_1 \\
m[1/2 + (i-1)] & x = x_i,\quad i = 1,\ldots,N \\
im & x_i < x < x_{i+1}, \quad i = 1,\ldots,N-1\\
M & x > x_N
\end{cases}
\end{equation}
where we have used the convention that integrating up to a $\delta$-function yields half the mass of integrating through it. To establish a correspondence between the discrete and continuum problems, we require that  $\psi_c(x_i) = \psi_d(x_i)$, which in turn determines the point-mass positions $x_i$. As $N \to \infty$ for fixed $M$, this step function $\psi_d$ converges uniformly to $\psi_c$. The correspondence goes in the opposite direction as well. If we begin with an ensemble of $\delta$-functions $\rho_d$, we can find the corresponding cumulative density $\psi_d$, interpolate to approximate $\psi_c$, and differentiate to find an approximate continuous density $\rho_c$.

With this correspondence established, we now describe our numerical method. Given an initial condition $\rho_c(x,0)$, we determine the corresponding discrete density $\rho_d(x,0)$ and the initial point-mass positions $x_i(0)$ as described above. Substituting (\ref{eq:rhod}) into the governing equations (\ref{eqnOfMotion}) yields a system of $N$ ordinary differential equations
\begin{equation}
\label{odes}
  \frac{dx_i}{dt} = \sum_{j \neq i} m_j f_s(x_i (t)- x_j(t)).
\end{equation}
See \cite{mkbs2003} for an introduction to discrete swarming models of this type.  We then solve the differential equations (\ref{odes}) numerically. From the new point-mass positions $x_i(t)$, we then reconstruct $\psi_d(x)$, $\psi_c(x)$, and $\rho_c(x,t)$, again using the correspondence described in the preceding paragraph.

Our numerical scheme has three sources of error. First, there is error associated with the integration of the ordinary differential equations (\ref{odes}). This error is easily controlled. We use the Matlab routine ODE45 for the numerical solution. The second source of error is interpolation error in the construction of $\rho_c$ from the location of the point masses. We interpolate $\psi_d$ to construct $\psi_c$ and then differentiate the polynomial analytically to obtain $\rho_c$. The  error is $\mathcal{O}(N^{-2})$. The third source of error comes from the approximation of the integral in the velocity term (\ref{velocityEquation}). We perform this quadrature using the point-masses for collocation, with an error of  $\mathcal{O}(N^{-2})$. For a full description of the numerical method and further details of the error analysis, see \cite{l2008}.

\bibliographystyle{siam.bst}
\bibliography{bibliography}

\begin{thebibliography}{10}

\bibitem{bt2008}
{\sc A.~J. Bernoff and C.~M. Topaz}, {\em Equilibrium configurations of
  interacting particles in one dimension}.
\newblock In prep., 2008.

\bibitem{bl2007}
{\sc A.~L. Bertozzi and T.~Laurent}, {\em Finite-time blow-up of solutions of
  an aggregation equation in $\mathbb{R}^n$}, Comm. Math. Phys., 274 (2001),
  pp.~717--735.

\bibitem{bv2005}
{\sc M.~Bodnar and J.~J.L. Velasquez}, {\em Derivation of macroscopic equations
  for individual cell-based models: A formal approach}, Math. Meth. Appl. Sci.,
  28 (2005), pp.~1757--1779.

\bibitem{bv2006}
\leavevmode\vrule height 2pt depth -1.6pt width 23pt, {\em An
  integro-differential equation arising as a limit of individual cell-based
  models}, J. Diff. Eq., 222 (2006), pp.~341--380.

\bibitem{bdt1999}
{\sc E.~Bonabeu, M.~Dorigo, and G.~Theraulaz}, {\em Swarm Intelligence: From
  Natural to Artificial Systems}, Santa Fe Institute Studies in the Sciences of
  Complexity, Oxford University Press, New York, 1999.

\bibitem{cdmbc2007}
{\sc Y.~L. Chuang, M.~R. D'Orsogna, D.~Marthaler, A.~L. Bertozzi, and L.~S.
  Chayes}, {\em State transitions and the continuum limit for a 2d interacting,
  self-propelled particle system}, Physica D, 232 (2007), pp.~33--47.

\bibitem{ckjrf2002}
{\sc I.~D. Couzin, J.~Krause, R.~James, G.~D. Ruxton, and N.~R. Franks}, {\em
  Collective memory and spatial sorting in animal groups}, J. Theor. Biol., 218
  (2002), pp.~1--11.

\bibitem{dcbc2006}
{\sc M.~R. D'Orsogna, Y.~L. Chuang, A.~L. Bertozzi, and L.~Chayes}, {\em
  Self-propelled particles with soft-core interactions: Patterns, stability,
  and collapse}, Phys. Rev. Lett., 96 (2006), pp.~104302.1--104302.4.

\bibitem{edl2007}
{\sc R.~Eftimie, G.~de~Vries, and M.~A. Lewis}, {\em Complex spatial group
  patters result from different animal communication mechanisms}, Proceedings
  of the National Academy of Sciences, 104 (2007), pp.~6974--6979.

\bibitem{edll2006}
{\sc R.~Eftimie, G.~de~Vries, M.~A. Lewis, and F.~Lutscher}, {\em Modeling
  group formation and activity patterns in self-organizing collectives of
  individuals}, Bulletin of Mathematical Biology, 69 (2007), pp.~1537--1565.

\bibitem{glr1996}
{\sc S.~Gueron, S.A. Levin, and D.I. Rubenstein}, {\em The dynamics of herds:
  From individuals to aggregations}, Journal of Theoretical Biology, 182
  (1996), pp.~85--98.

\bibitem{l2008}
{\sc A.~J. Leverentz}, {\em An integrodifferential equation modeling 1-d
  swarming behavior}.
\newblock Senior thesis, Harvey Mudd College, 2008.

\bibitem{lrc2001}
{\sc H.~Levine, W.~J. Rappel, and I.~Cohen}, {\em Self-organization in systems
  of self-propelled particles}, Phys. Rev. E, 63 (2001),
  pp.~017101.1--017101.4.

\bibitem{mk1999}
{\sc A.~Mogilner and L.~Edelstein-Keshet}, {\em A non-local model for a swarm},
  J. Math. Bio., 38 (1999), pp.~534--570.

\bibitem{mkbs2003}
{\sc A.~Mogilner, L.~Edelstein-Keshet, L.~Bent, and A.~Spiros}, {\em Mutual
  interactions, potentials, and individual distance in a social aggregation},
  J. Math. Bio., 47 (2003), pp.~353--389.

\bibitem{ogk2001}
{\sc A.~Okubo, D.~Gr{\"{u}}nbaum, and L.~Edelstein-Keshet}, {\em The dynamics
  of animal grouping}, in Diffusion and Ecological Problems, A.~Okubo and S.~A.
  Levin, eds., vol.~14 of Interdisciplinary Applied Mathematics: Mathematical
  Biology, Springer, New York, second~ed., 2001, ch.~7, pp.~197--237.

\bibitem{pk1999}
{\sc J.~K. Parrish and L.~Edelstein-Keshet}, {\em Complexity, pattern, and
  evolutionary trade-offs in animal aggregation}, Science, 284 (1999),
  pp.~99--101.

\bibitem{pm2005}
{\sc S.~R. Partan and P.~Marler}, {\em Issues in the classification of
  multimodal communication signals}, Am. Nat., 166 (2005), pp.~231--245.

\bibitem{p2005}
{\sc K.~M. Passino}, {\em Biomimicry for Optimization, Control, and
  Automation}, Springer, London, 2005.

\bibitem{tk1998}
{\sc D.~Tilman and P.~Kareiva}, eds., {\em Spatial Ecology: The Role of Space
  in Population Dynamics and Interspecific Interactions}, Princeton University
  Press, Princeton, NJ, 1998.

\bibitem{tblt2008}
{\sc C.~M. Topaz, A.~J. Bernoff, S.~Logan, and W.~Toolson}, {\em A model for
  rolling swarms of locusts}, Euro. Phys. J. ST, 157 (2008), pp.~93--109.

\bibitem{tb2004}
{\sc C.~M. Topaz and A.~L. Bertozzi}, {\em Swarming patterns in a
  two-dimensional kinematic model for biological groups}, SIAM J. Appl. Math.,
  65 (2004), pp.~152--174.

\bibitem{tbl2006}
{\sc C.~M. Topaz, A.~L. Bertozzi, and M.~A. Lewis}, {\em A nonlocal continuum
  model for biological aggregation}, Bull. Math. Bio., 68 (2006),
  pp.~1601--1623.

\bibitem{w1974}
{\sc G.~B. Whitham}, {\em Linear and Nonlinear Waves}, Wiley, New York, 1974.

\bibitem{wb1998}
{\sc T.~P. Witelski and A.~J. Bernoff}, {\em Self-similar asymptotics for
  linear and nonlinear diffusion equations}, Stud. Appl. Math., 100 (1998),
  pp.~153--193.

\bibitem{zb1958}
{\sc Y.~B. Zeldovich and G.~I. Barenblatt}, {\em Asymptotic properties of
  self-preserving solutions of equations of unsteady motion of gas through
  porous media}, Dokl. Akad. Nauk SSR, 118 (1958), pp.~671--674.

\end{thebibliography}

\end{document}